\documentclass{PoS}

\usepackage{amssymb,amsmath,amsfonts,amsthm,mathrsfs}
\usepackage{array,multirow}

\usepackage{graphicx}
\usepackage{epstopdf} 
\graphicspath{{./figures/}}

\usepackage{wrapfig}

\newcommand{\Fig}[1]{Fig.~\ref{#1}}

\newcommand{\Eq}[1]{Eq.\,\eqref{#1}}
\newcommand{\Ref}[1]{Ref.~\cite{#1}}

\newcommand{\Tr}{\operatorname{Tr}}

\title{
Quantum entanglement in SU(3) lattice Yang-Mills theory at zero and finite temperatures
}

\ShortTitle{Entanglement entropy}

\author{\speaker{Y.~Nakagawa}
\\
Graduate School of Science and Technology,
Niigata University, Niigata, 950-2181, Japan
\\
E-mail: \email{nakagawa@muse.sc.niigata-u.ac.jp}
}

\author{A.~Nakamura
\\
Research Institute for Information Science and Education,
Hiroshima University, Higashi-Hiroshima, Hiroshima, 739-8521, Japan
\\
E-mail: \email{nakamura@riise.hiroshima-u.ac.jp}
}
\author{S.~Motoki
\\
Graduate School of Bio-Sphere Science,
Hiroshima University, Higashi-Hiroshima, Hiroshima, 739-8521, Japan
\\
E-mail: \email{motoki-shinji@hiroshima-u.ac.jp}
}
\author{V.I. Zakharov
\\
ITEP, B. Cheremushkinskaya 25, Moscow, 117218, Russia \\
Max-Planck-Institut f\"ur Physik, F\"ohringer Ring 6, 80805 M\"unich, Germany
\\
E-mail: \email{xxz@mppmu.mpg.de}
}

\abstract{
We examine the entanglement properties of the Yang-Mills theory
by calculating $\alpha$ entanglement entropy with $\alpha=2$
using a SU(3) quenched lattice gauge simulation both in
the confinement and the deconfinement phases.
In the confinement phase, the derivative of the $\alpha$ entropy 
with respect to the size $l$ of the subregion, whose entanglement
properties are interested in, scales as $1/l^3$,
and a clear discontinuity cannot be found within our statistical errors.
The $\alpha$ entropy in the deconfinement phase saturates at large $l$.
The saturation value is comparable with the thermal entropy of
the pure Yang-Mills theory, indicating that the $\alpha$ entropy obeys
the volume law at large $l$ in the deconfinement phase.
}

\FullConference{The XXVIII International Symposium on Lattice Field Theory\\
		 July 14-19 2010\\
		 Villasimius, Italy}

\begin{document}

\section{Introduction}
\label{sec:Introduction}

Various quantum systems show entanglement properties and
they receive much attention in quantum information theory
and condensed matter physics.
Entanglement entropy is one of quantities measuring quantum entanglement.
A typical example of the entangled state in quantum mechanical systems
is a two spin-1/2 system in spin singlet state, which is widely used
to discuss the EPR paradox, one of major topics in quantum physics.
Entanglement entropy can be defined not only in quantum mechanical systems
but also in quantum field theories.

In quantum mechanical systems,
fundamental degrees of freedom are particles and
quantum entanglement measures how much two or more particles are
quantum mechanically correlated with each others.
In quantum field theories, we focus on quantum entanglement of two
or more sibregions.
The entanglement entropy between two subregions,
a subregion $A$ of size $l$ and its complement $B$,
measures how the spatial subregion in a total system
is entangled quantum mechanically with its complement.

Quantum entanglement of ground states has been widely discussed
in condensed matter physics
(for a review, see \cite{Amico:2007ag}).
For example, the entanglement entropy in the Ising chain model
shows a divergent behavior at the critical point
while it saturates in the non-critical regime.
It means that the entanglement entropy serves as an order parameter
of quantum phase transitions.
Therefore, the entanglement entropy is a useful quantity to investigate
phase structures of quantum systems.

\begin{figure}[tb]
\begin{minipage}{0.45\hsize}
\begin{center}
\resizebox{1.\textwidth}{!}
{\includegraphics{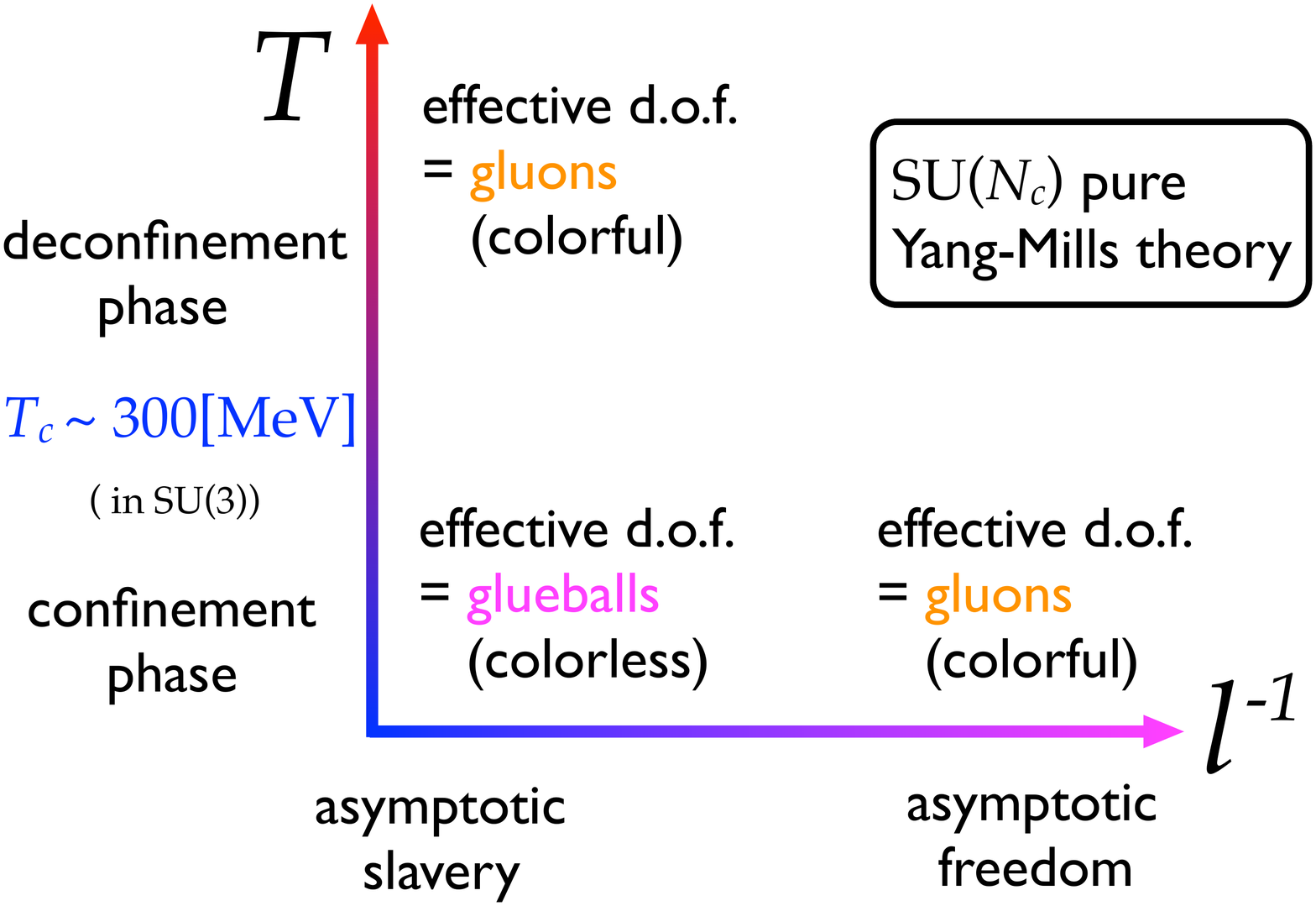}}
\caption{Schematic picture of the pure Yang-Mill system.}
\label{fig:motivation}
\end{center}
\end{minipage}
\hspace{0.05\hsize}
\begin{minipage}{0.45\hsize}
\begin{center}
\resizebox{0.85\textwidth}{!} 
{\includegraphics{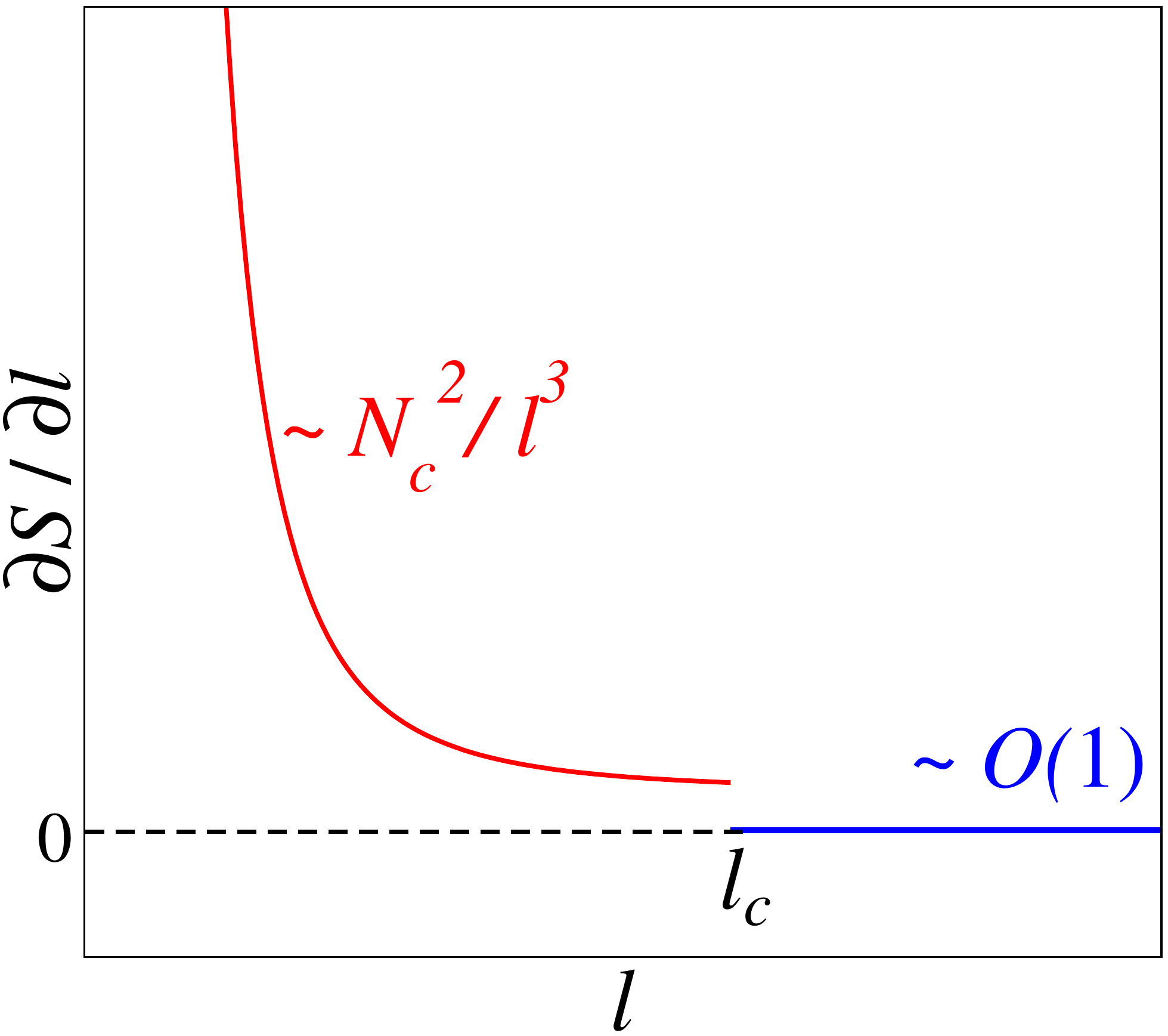}}
\caption{
A typical example of the holograhic prediction showing
the discontinuity of the entanglement entropy $S_A$.
}
\label{fig:dSdl_AdS_CFT}
\end{center}
\end{minipage}
\end{figure}

The entanglement entropy of the pure Yang-Mills theory is particularly interesting.
A schematic picture of the pure Yang-Mills system is drawn in
\Fig{fig:motivation}.
The Yang-Mills theory is an asymptotically free theory
and the high energy phenomena in QCD can well be described by
gluon and quark degrees of freedom using the perturbation theory.
At low energies, on the other hand, color degrees of freedom are confined
in hadrons and the Yang-Mills system is described by the colorless hadrons.
This may remind us of the deconfinement phase transition;
the color degrees of freedom are released above the critical temperature,
and gluons (and quarks in QCD) play a major role as effective degrees of freedom
while those in the confinement phase are glueballs (or hadrons).
Thus, one might ask if there is a critical distance scale
at which the effective degrees of freedom change from colorful objects
to colorless objects as the critical temperature of the deconfinement phase transition.

Recently, gauge/gravity correspondence has been extensively studied
and it provides a powerful tool to study non-perturbative infrared dynamics
of confining gauge theories.
Beginning with the pioneering work by Ryu and Takayanagi
\cite{Ryu:2006bv},
the holographic approach is applied to the calculation of the entanglement entropy
(for a review on the holographic calculation, see
\cite{Nishioka:2009un}).
In this approach, the entanglement entropy of gauge theories is obtained
by calculating geodesics (minimal surface bending down to the bulk space)
in the gravity side, similar to the calculation of the Wilson loop
in the holographic approach.
The boundary of geodesics coincides the boundary of partitioned subsystems
in gauge theory side.
The entanglement entropy has been studied for various confining backgrounds
\cite{Nishioka:2006gr,Klebanov:2007ws}.
It has been argued that the entanglement entropy could exhibit a non-analytic
behavior with respect to the size $l$ of the subregion;
an $O(N_c^2)$ solution dominates at small $l$, and a $l$-independent $O(1)$ solution
dominates above some critical length $l^{\ast}$
(see \Fig{fig:dSdl_AdS_CFT}).
This indicates that the effective degrees of freedom change from colorful objects
to colorless objects, and the critical length $l^{\ast}$ plays a role of
the inverse of the critical temperature of the deconfinement phase transition.

The entanglement entropy in SU(2) lattice gauge theory has been studied
by Velytsky
\cite{Velytsky:2008rs}
and Bividovich and Polikarpov
\cite{Buividovich:2008kq}.
In \Ref{Velytsky:2008rs}, SU(N) lattice gauge theories are
studied in Migdal-Kadanoff approximation, 
and in \Ref{Buividovich:2008kq}, SU(2) lattice gauge theory
is numerically investigated, and there is an indication
that the derivative of the entanglement entropy shows a discontinuous
change at some critical length scale $l^{\ast}$ and it vanishes at large $l$.

In this study, we investigate $\alpha$ entanglement entropy
in SU(3) pure Yang-Mills theory using lattice Monte Carlo simulations.
Instead of directly calculating the entropy, we adopt numerical technique
to evaluate the entanglement entropy, which has also been used in
\cite{Buividovich:2008kq}
(originally proposed in
\cite{Endrodi:2007tq,Fodor:2007sy}
in order to calculate the pressure in the deconfined phase).

\section{Entanglement entropy}
\label{sec:definition}

\begin{wrapfigure}{l}{5.5cm}
\begin{center}
\vspace{-0.5cm}
\resizebox{0.3\textwidth}{!} 
{\includegraphics{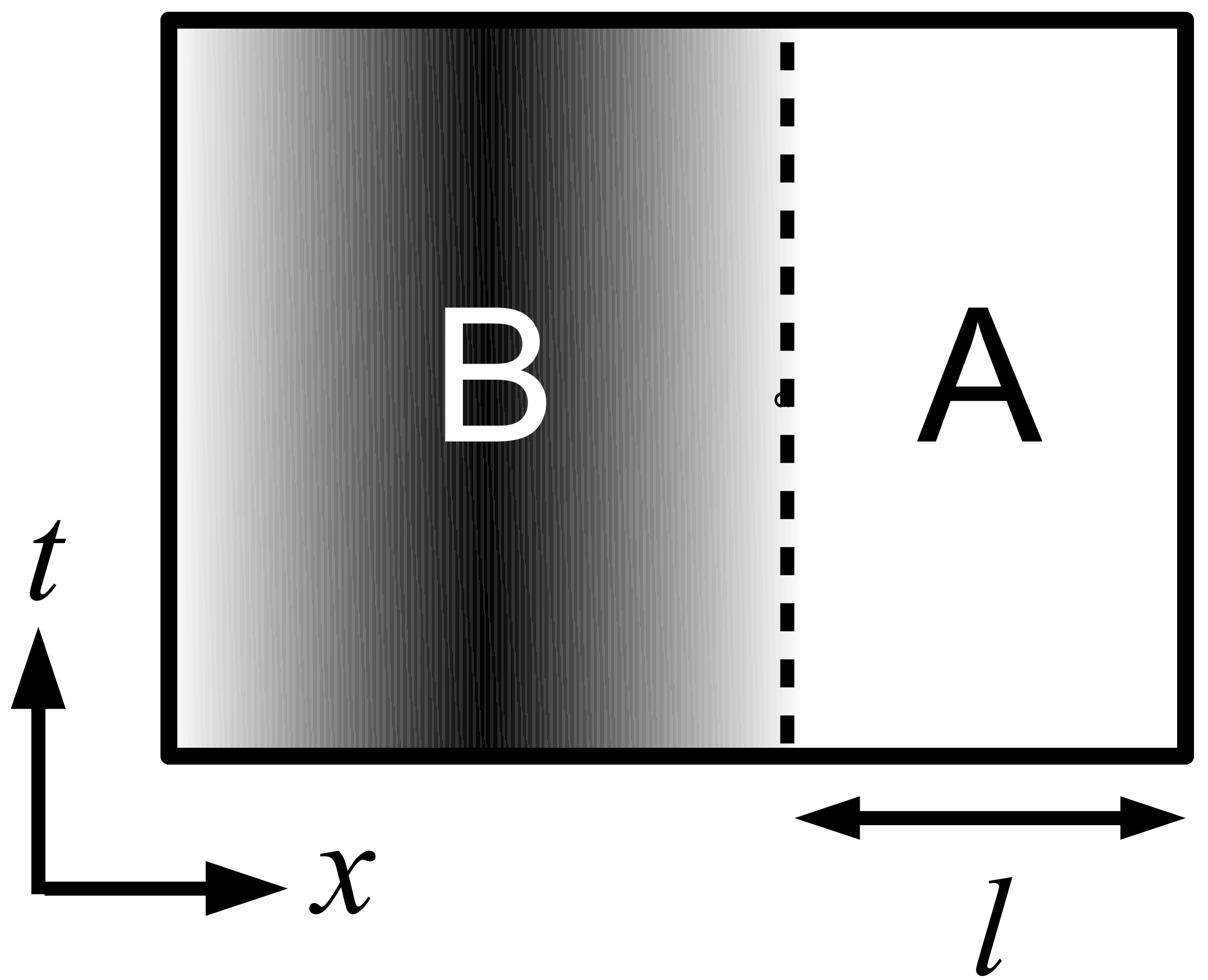}}
\caption{
The complementary regions $A$ and $B$
separated by an imaginary boundary at $x=l$.
}
\label{fig:two_regions_AB}
\end{center}
\end{wrapfigure}

Entanglement entropy of a pure state $|\Psi\rangle$
is defined as follows.
We divide the total system into subregion $A$ and its complement $B$
(see \Fig{fig:two_regions_AB}).
Let $l$ be the size of the system $A$ in the $x$ direction.
The density matrix of the system is
$
\rho = | \Psi \rangle \langle \Psi |.
$
At zero temperature, the ground state is a pure state
and the von Neumann entropy of the system is zero.
The reduced density matrix obtained by tracing out
the degrees of freedom in the region $B$,
\begin{equation}
\rho_A = \Tr_B \rho = \Tr_B | \Psi \rangle \langle \Psi |,
\end{equation}
describes the density matrix for an observer who can only access
to the subregion $A$.
Although we start off with a pure state with vanishing von Neumann entropy,
the state corresponding to the reduced density matrix is generally a mixed state.
$\rho_A$ contains the information on the quantum degrees of
freedom traced out.
The entanglement entropy is defined as the von Neumann entropy
of the reduced density matrix,
\begin{equation}\label{eq:entanglementE}
S_A = - \Tr \rho_A \ln \rho_A.
\end{equation}
Some properties of the entanglement entropy can be found in
\cite{Nielsen:2000fk}.

\section{Replica method}
\label{sec:replica}

\begin{wrapfigure}{l}{4cm}
\begin{center}
\vspace{-1cm}
\resizebox{0.25\textwidth}{!} 
{\includegraphics{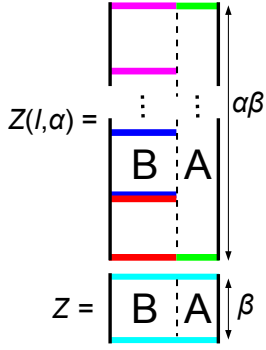}}
\caption{
Schematic picture for the system with $\alpha$ cuts
in $x-t$ plane.
In the region $A$ ($B$), the periodic boundary condition is imposed
with the period $\alpha\beta$ ($\beta$).
}
\label{fig:replica_trick}
\end{center}
\end{wrapfigure}

In order to evaluate the entanglement entropy,
we apply the replica trick.
The detail of the derivation is given in
\cite{Calabrese:2004eu}.
The point is that the entanglement entropy defined in
\Eq{eq:entanglementE} can be represented in the form,
$S_A = - \lim_{\alpha \to 1} \partial/\partial\alpha \ln \Tr_A \rho_A^\alpha$.
The trace of the $\alpha$-th power of the reduced density matrix $\rho_A$
is given by the ratio of the partition functions,
\begin{equation}
\Tr \rho_A^\alpha = \frac{Z(l,\alpha)}{Z^\alpha}.
\end{equation}
Here $Z(l,\alpha)$ is the partition function of the system having
special topology, the $\alpha$-sheeted Riemann surface,
and $Z=Z(\alpha=1)$.
The field variables in the region $A$ is periodically identified
with the interval $\alpha\beta$ ($\beta=1/T$ is the lattice size
in the temporal direction) while in the region $B$
the periodic boundary condition is imposed with the period $\beta$
(see \Fig{fig:replica_trick}).

The entanglement entropy is then given by
\begin{equation}
S_A(l) = - \lim_{\alpha \to 1} \frac{\partial}{\partial \alpha}
\ln \left( \frac{Z(l,\alpha)}{Z^\alpha} \right).
\end{equation}
The derivative of $S_A(l)$ with respect to $l$, which is
free of the ultraviolet divergence, can be expressed as follows;
\begin{equation}\label{eq:dSdl}
\frac{\partial S_A(l)}{dl}
= \frac{\partial}{\partial l} \left[ - \lim_{\alpha \to 1} 
\frac{\partial}{\partial \alpha} \ln \left( \frac{Z(l,\alpha)}{Z^\alpha}
\right) \right]
= \lim_{\alpha \to 1}
\frac{\partial}{\partial l} \frac{\partial}{\partial \alpha} F[l,\alpha].
\end{equation}
That is, in order to calculate $\partial S_A/\partial l$,
we evaluate the free energy of the system having $\alpha$ cuts
as is depicted in \Fig{fig:replica_trick}, take the derivative
with respect to $\alpha$ and $l$, and then take the limit $\alpha \to 1$.
Thus, the evaluation of the entanglement entropy is reduced
to the calculation the free energy of the system with $\alpha$ cuts.

\section{Lattice setup and observables}
\label{sec:lattice_setup}
\vspace{-0.3cm}

In the lattice simulations, the derivative in \Eq{eq:dSdl}
is replaced by the finite difference,
and we estimate the derivative by
\begin{equation}\label{eq:dSdl_lat}
\lim_{\alpha \to 1} \frac{\partial}{\partial l}
\frac{\partial}{\partial \alpha} F[A,\alpha]
 \to 
\frac{\partial}{\partial l}
\lim_{\alpha \to 1} \left( F[l,\alpha+1] - F[l,\alpha] \right)
 \to  \frac{F[l+a,\alpha=2] - F[l,\alpha=2]}{a}.
\end{equation}
We note that $\partial F[l,\alpha=1]/\partial l$ drops out
since $F[l,\alpha=1]$ does not depend on $l$.
The difference of the free energies can be evaluated numerically
by introducing an `interpolating action' which interpolates
two actions corresponding to two free energies
\cite{Endrodi:2007tq,Fodor:2007sy},
$S_{\textrm{int}} = (1-\gamma) S_l[U] + \gamma S_{l+a}[U]$.
$S_l$ and $S_{l+a}$ represent the actions
corresponding to $F[l,\alpha=2]$ and $F[l+a,\alpha=2]$ in \Eq{eq:dSdl_lat}.
It is easy to show that
\begin{equation}\label{eq:alpha_integral}
  F[l+a,\alpha=2] - F[l,\alpha=2]
  = - \int^1_0 d\gamma \frac{\partial}{\partial \gamma} \ln Z(l,\gamma)
  = \int^1_0 d \gamma \left\langle S_{l+a}[U] - S_l[U] \right\rangle_{\gamma}.
\end{equation}
Here $\langle \cdot \rangle_{\gamma}$ refers to the Monte Carlo average
with the interpolating action $S_{\textrm{int}}$.
Therefore, the $\alpha=2$ entanglement entropy can be evaluated numerically
by updating gauge configurations with the interpolating action
on the lattice with $\alpha=2$ cuts,
calculating the action differences for various $\gamma$,
and performing a numerical integration over $\gamma$.
In order to evaluate the integral in \Eq{eq:alpha_integral},
we calculated the action differences from $\gamma=0$ to 1
by the step 0.1, and employed the Simpson's rule
to evaluate the integration numerically, which interpolates
neighboring points by a quadratic curve.

We adopt the heat-bath Monte Carlo technique with the standard
plaquette action to generate lattice configurations.
First 5000 sweeps are discarded for thermalization,
and the measurement has been done every 100 sweeps.
The number of configurations for each $\beta$ and lattice
size is about 3000 to 8000.
The statistical errors are estimated by the jackknife method.

\section{Simulation results}
\label{sec:results_dSdl}

\subsection{$\alpha$ entanglement entropy at zero temperature}

\begin{figure}[b]
\begin{minipage}{0.47\hsize}\begin{center}
\resizebox{1.0\textwidth}{!}{\includegraphics
{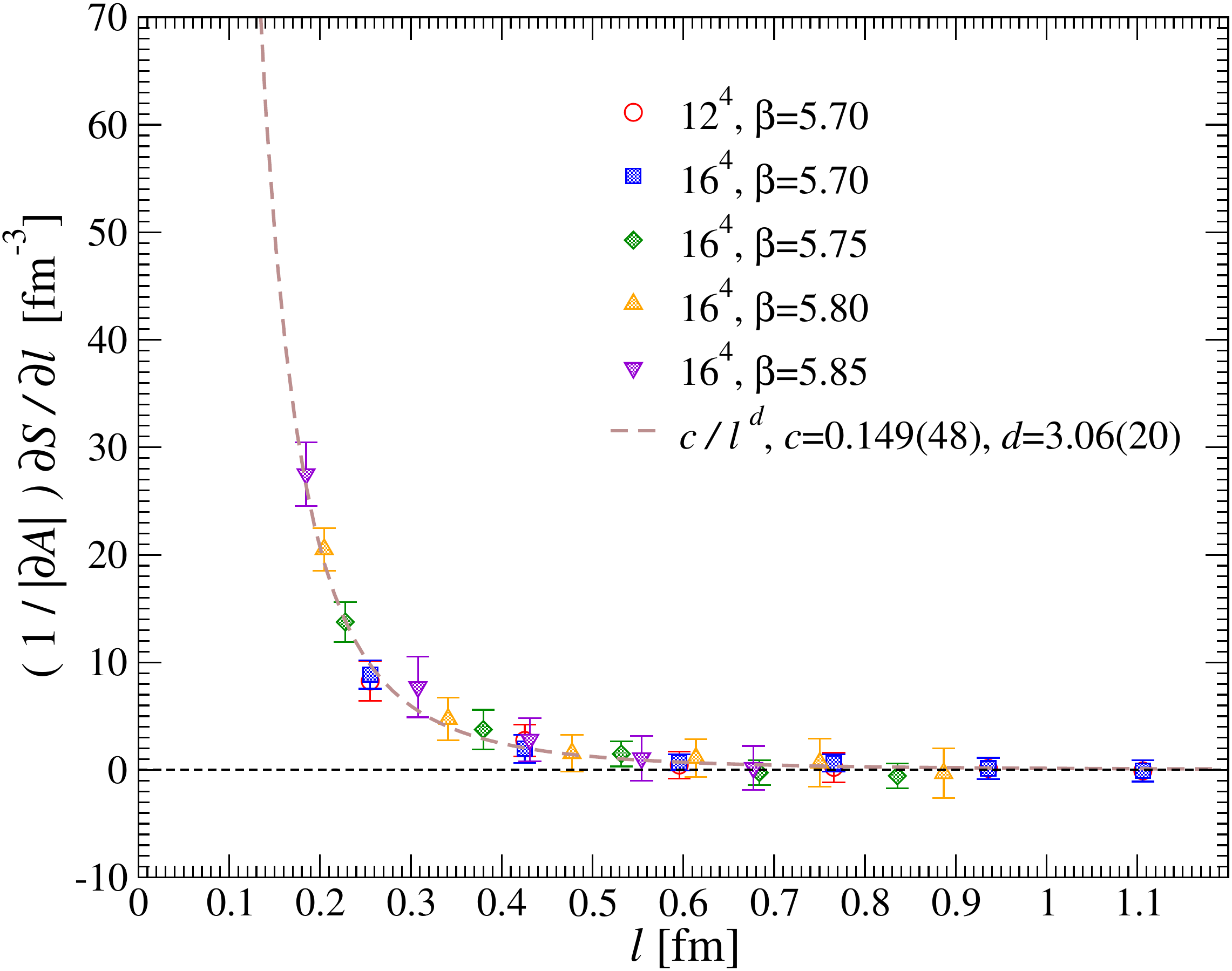}
}
\end{center}\end{minipage}
\hspace{0.03\hsize}
\begin{minipage}{0.47\hsize}\begin{center}
\resizebox{1.0\textwidth}{!}{\includegraphics
{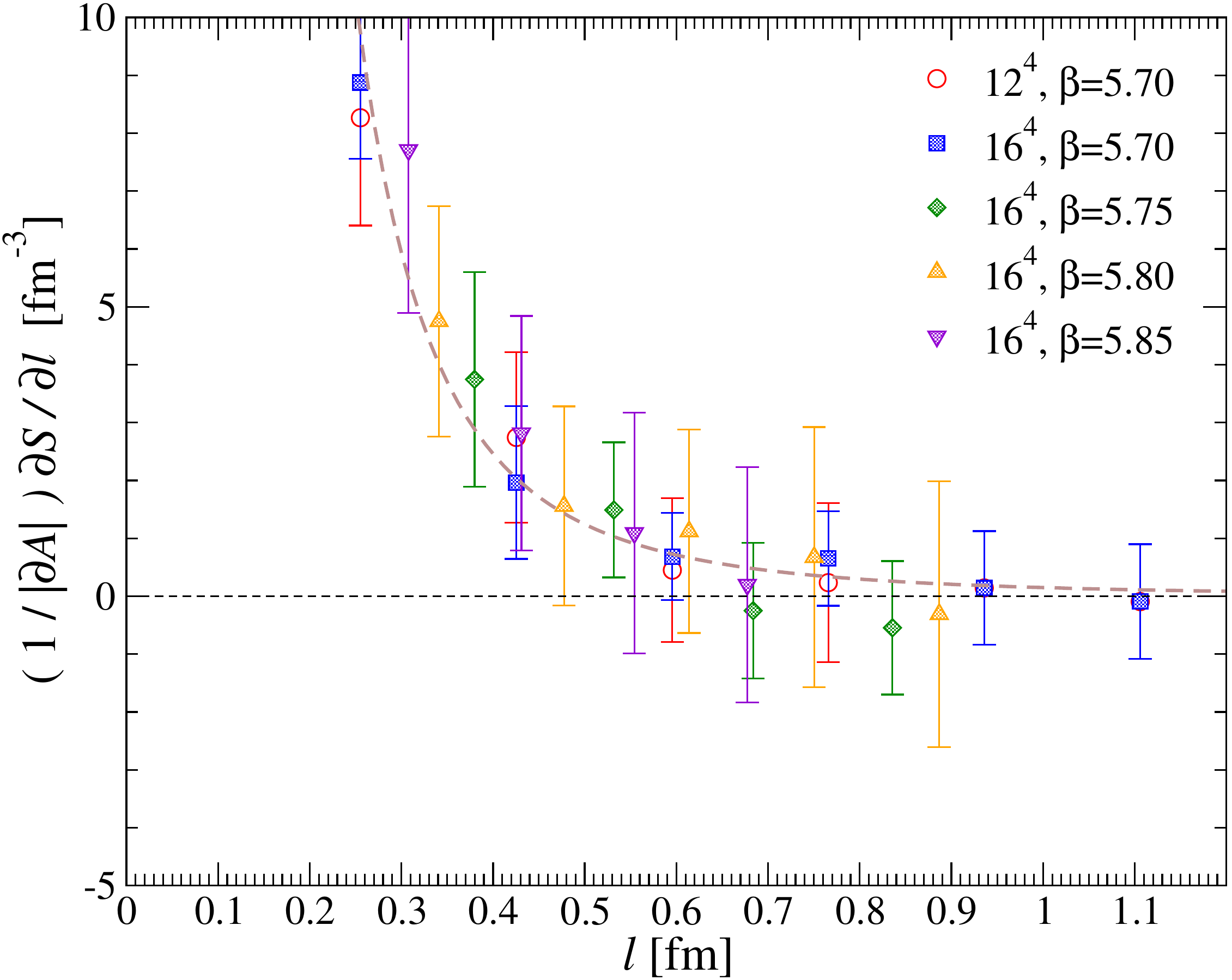}
}
\end{center}\end{minipage}
\caption{
$(1/|\partial A|)\partial S_A/\partial l$
in the confinement phase.
The dashed curve is the fit of the data by the function
$c/l^d$ with the fitted values $c=0.149(48), d=3.06(20)$.
The right panel shows the zoom up of the left panel
to make near-zero region more visible.
}
\label{fig:dSdl_SU3}
\end{figure}

The derivative of $S_A(l)$ with respect to $l$
in the confinement phase is plotted in \Fig{fig:dSdl_SU3}.
$\partial S_A(l)/\partial l$ is normalized by the area
of the common boundary of the two subregions, $|\partial A|$.
We observe that data on $12^4$ and $16^4$ agree within statistical errors.
This implies that the derivative of the $\alpha=2$ entanglement entropy
is proportional to the area of the boundary,
namely, entanglement entropy obeys an area law at zero temperature.

In the small $l$ region, the $\alpha=2$ entanglement entropy is
expected to scale as $1/l^2$ from the dimensional analysis.
That is, $\partial S_A/\partial l$ behaves as $1/l^3$ at small $l$.
This behavior is exactly what the entanglement entropy
in conformal field theory in (3+1)-dimensional spacetime shows.
In order to confirm this, we fitted data with the function
$\partial S_A/\partial l = c \left( 1/l \right)^{d}$,
and we obtain $c=0.149(48), d=3.06(20), \chi^2/ndf = 0.192$.
The fitted function is plotted in \Fig{fig:dSdl_SU3}
by the dashed curve.
Our result does not show a clear sign of the discontinuity
in $\partial S_A/\partial l$.
Since the derivative of $S_A$ rapidly decreases,
the signal-to-noise ratio becomes quite small at large $l$ and
it is very difficult to locate the critical length
of the entanglement entropy numerically, even if it exists.
It can be safely stated that our results exclude the possibility
of the existence of the critical length at below 0.4 [fm].

\subsection{$\alpha$ entanglement entropy below and above the critical temperature}

\begin{figure}[b]
\begin{minipage}{0.47\hsize}\begin{center}
\resizebox{1.0\textwidth}{!}
{\includegraphics{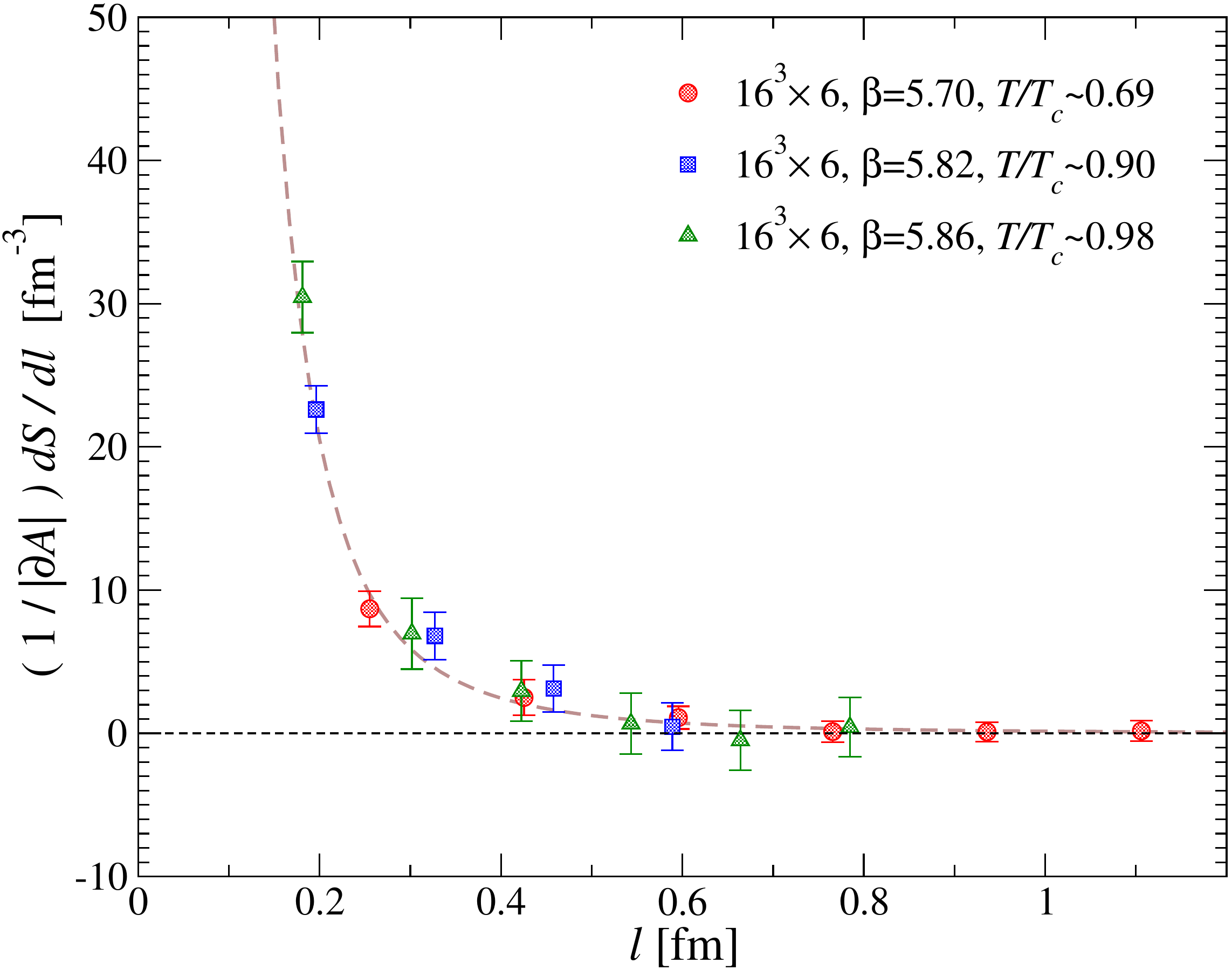}}
\end{center}\end{minipage}
\hspace{0.03\hsize}
\begin{minipage}{0.47\hsize}\begin{center}
\resizebox{1.0\textwidth}{!}
{\includegraphics{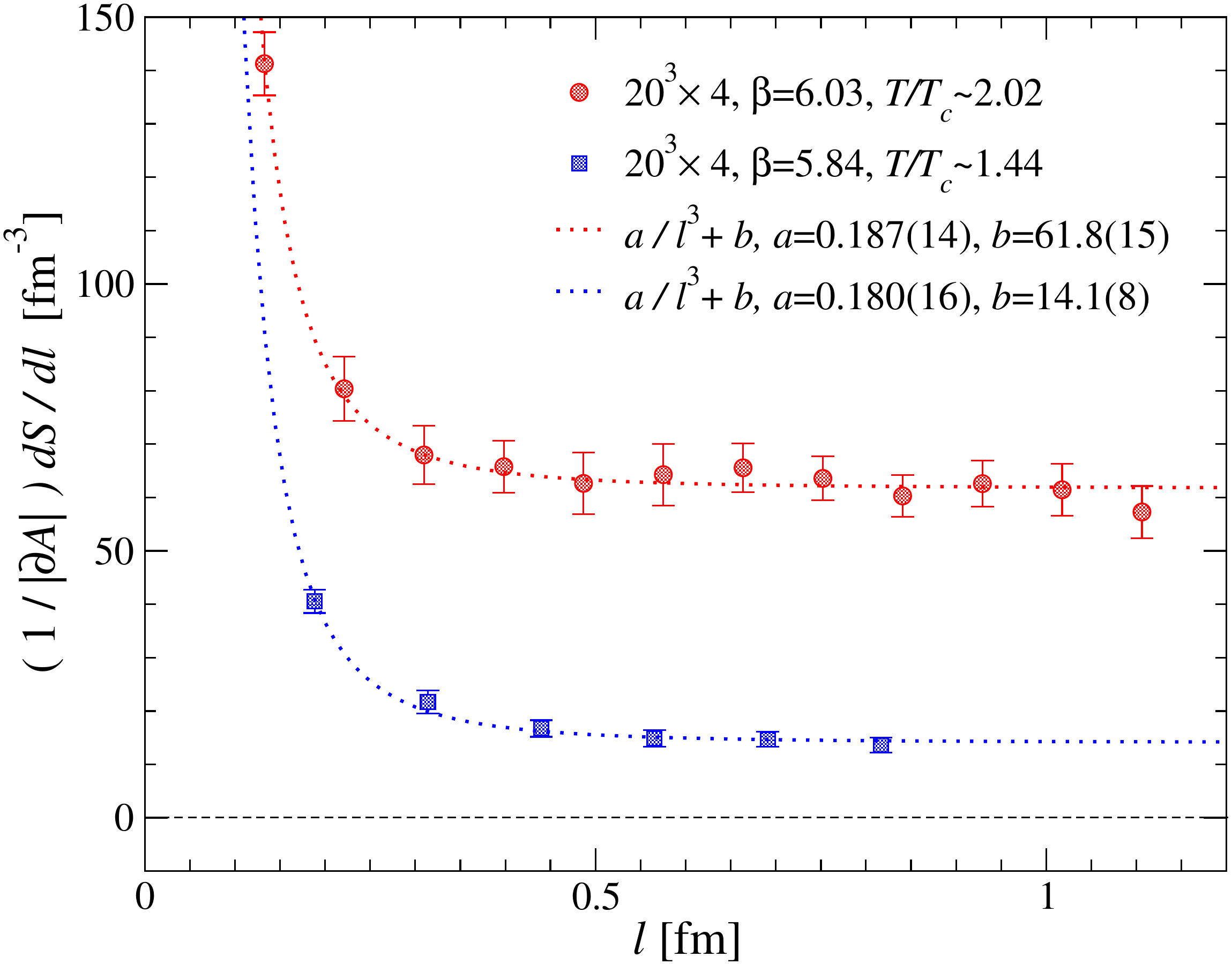}}
\end{center}\end{minipage}
\caption{
Left panel: 
$(1/|\partial A|)\partial S_A/\partial l$ below the critical temperature.
The dashed curve is the fitted function at zero temperature.
Right panel:
$(1/|\partial A|)\partial S_A/\partial l$ above the critical temperature.
The dotted curves show the fits of the data by the function $c/l^3+d$.
}
\label{fig:dSdl_SU3_finiteTs}
\end{figure}

The left panel of \Fig{fig:dSdl_SU3_finiteTs} shows
the derivative of the $\alpha$ entanglement entropy
below the critical temperature.
The fitted function at zero temperature,
$(1/|\partial A|) \partial S_A/\partial l = 0.149/l^{3.06}$,
is drawn by the dashed curve.
We observe that the data agree with the fitted function
of the zero temperature result.
This indicates that the $\alpha$ entanglement entropy does not show
a clear temperature dependence below the critical temperature.

The numerical result above the critical temperature
is given in the right panel of \Fig{fig:dSdl_SU3_finiteTs}.
We see that $\partial S_A/\partial l$ does not approach zero
but saturates at large $l$.
At zero temperature, the ground state is a pure state
and the von Neumann entropy is zero.
By contrast, the ground state at finite temperature is a mixed state
and the von Neumann entropy (thermal entropy) takes a finite value.
This means that at finite temperature,
the entanglement entropy measures not only
the quantum mechanical correlation between the two spatial subregions
but also the thermal entropy of the subregion.
Since the thermal entropy of the SU(3) Yang-Mills theory
rapidly increases in the vicinity of the critical temperature,
the saturation value of the entanglement entropy may be considered
as the thermal entropy of the subregion $A$.
We note that the asymptotic behavior of $(1/|\partial A|)\partial S_A/\partial l$
implies that the entanglement entropy obeys the volume law at large $l$
above the critical temperature.

We fitted the data with the function $a/l^3+b$,
and we obtain
\begin{equation}
\begin{array}{lll}
a = 0.180(16), & ~~ b = 14.1(8)  & ~~ (T/T_c \sim 1.44) \\
a = 0.187(14), & ~~ b = 61.8(15) & ~~ (T/T_c \sim 2.02).
\end{array}
\end{equation}
We note that the coefficient of the $1/l^3$ term agrees with each other.
In order to compare the asymptotic value of $\partial S_A/\partial l$
to the thermal entropy, we estimated the thermal entropy
at $T/T_c\sim 1.44$ and $2.02$ by reading the values off the figure in
\Ref{Boyd:1996bx}.
A rough estimate gives
\begin{equation}
\begin{array}{ll}
s = 17 & ~~ (T/T_c \sim 1.44) \\
s = 56 & ~~ (T/T_c \sim 2.02).
\end{array}
\end{equation}
These are comparable with the asymptotic values of $\partial S_A/\partial l$.

\section{Summary and conclusion}
\label{sec:summary}
\vspace{-0.3cm}

We studied the $\alpha$ entanglement entropy of the Yang-Mills vacuum
with $\alpha=2$ using lattice Monte Carlo simulations.
The entanglement entropy measures the quantum correlation
between spatial subregions.
We find that at zero temperature
the derivative of the $\alpha=2$ entropy 
with respect to $l$ is well fitted by the function
$c/l^d$ with d=3.06(20).
The exponent $d$ is consistent with that in the conformal field theory.
A clear discontinuity in $\partial S_A/\partial l$
was not observed within the statistical errors,
which is arguded in the models of the gauge/gravity correspondence.
Furthermore, we observe that the $\alpha$ entropy is
almost temperature independent below the critical temperature.
Above the critical temperature, $\alpha=2$ entropy
does not approach zero but saturates at large $l$.
Since the ground state of the finite temperature system is
a mixed state, this implies that the entanglement entropy
measures not only the correlation between the spatial subregions
but also the thermal entropy of the subregion, which dominates
at large $l$.
Indeed, our fitted result of the asymptotic values is
comparable with the thermal entropy of the pure SU(3) Yang-Mills theory.

\vspace{0.3cm}
\noindent
{\large\bf Acknowledgements}

The simulation was performed on
NEC SX-8R at RCNP, Osaka University
and NEC SX-9 at CMC, Osaka University.
The work is partially supported by
Grant-in-Aid for Scientific Research by
Monbu-kagakusyo, No. 20340055.

\providecommand{\href}[2]{#2}\begingroup\raggedright\endgroup


\begin{thebibliography}{10}
\vspace{-0.2cm}
\bibitem{Amico:2007ag}
L.~Amico, R.~Fazio, A.~Osterloh, and V.~Vedral,
  \href{http://dx.doi.org/10.1103//RevModPhys.80.517}{{\em
  Rev.Mod.Phys.} {\bfseries 80} (2008) 517},
  \href{http://arxiv.org/abs/quant-ph/0703044}{{\ttfamily
  quant-ph/0703044}}.

\bibitem{Ryu:2006bv}
S.~Ryu and T.~Takayanagi,
  \href{http://dx.doi.org/10.1103/PhysRevLett.96.181602}{{\em
  Phys.Rev.Lett.} {\bfseries 96} (2006) 181602},
  \href{http://arxiv.org/abs/hep-th/0603001}{{\ttfamily hep-th/0603001}}.

\bibitem{Nishioka:2009un}
T.~Nishioka, S.~Ryu, and T.~Takayanagi,
  \href{http://dx.doi.org/10.1088/1751-8113/42/50/504008}{{\em J.
  Phys.} {\bfseries A42} (2009) 504008},
\href{http://arxiv.org/abs/0905.0932}{{\ttfamily arXiv:0905.0932}}.

\bibitem{Nishioka:2006gr}
T.~Nishioka and T.~Takayanagi,
  \href{http://dx.doi.org/10.1088/1126-6708/2007/01/090}{{\em
  JHEP} {\bfseries 0701} (2007) 090},
  \href{http://arxiv.org/abs/hep-th/0611035}{{\ttfamily hep-th/0611035}}.

\bibitem{Klebanov:2007ws}
I.~R. Klebanov, D.~Kutasov, and A.~Murugan,
  \href{http://dx.doi.org/10.1016/j.nuclphysb.2007.12.017}{{\em
  Nucl.Phys.} {\bfseries B796} (2008) 274},
  \href{http://arxiv.org/abs/0709.2140}{{\ttfamily arXiv:0709.2140}}.

\bibitem{Velytsky:2008rs}
A.~Velytsky,
  \href{http://dx.doi.org/10.1103/PhysRevD.77.085021}{{\em Phys. Rev.}
  {\bfseries D77} (2008) 085021},
\href{http://arxiv.org/abs/0801.4111}{{\ttfamily arXiv:0801.4111}}.

\bibitem{Buividovich:2008kq}
P.~V. Buividovich and M.~I. Polikarpov,
  \href{http://dx.doi.org/10.1016/j.nuclphysb.2008.04.024}{{\em Nucl. Phys.}
  {\bfseries B802} (2008) 458},
\href{http://arxiv.org/abs/0802.4247}{{\ttfamily arXiv:0802.4247}}.

\bibitem{Endrodi:2007tq}
G.~Endrodi, Z.~Fodor, S.~Katz, and K.~Szabo,
  {\em PoS} {\bfseries LAT2007} (2007) 228,
  \href{http://arxiv.org/abs/0710.4197}{{\ttfamily arXiv:0710.4197}}.

\bibitem{Fodor:2007sy}
Z.~Fodor,
  {\em PoS} {\bfseries LAT2007} (2007) 011,
  \href{http://arxiv.org/abs/0711.0336}{{\ttfamily arXiv:0711.0336}}.

\bibitem{Nielsen:2000fk}
M.~A. Nielsen and I.~L. Chuang, {\em Quantum Computation and Quantum
  Information}.
\newblock Cambridge University Press, 2000.

\bibitem{Calabrese:2004eu}
P.~Calabrese and J.~L. Cardy,
{\em J. Stat. Mech.} {\bfseries 0406} (2004) P002,
\href{http://arxiv.org/abs/hep-th/0405152}{{\ttfamily hep-th/0405152}}.

\bibitem{Boyd:1996bx}
G.~Boyd {\em et al.},
  \href{http://dx.doi.org/10.1016/0550-3213(96)00170-8}{{\em Nucl. Phys.}
  {\bfseries B469} (1996) 419},
\href{http://arxiv.org/abs/hep-lat/9602007}{{\ttfamily hep-lat/9602007}}.

\end{thebibliography}
\end{document}